\documentclass[10pt,aps,prx,twocolumn,superscriptaddress,showkeys]{revtex4-2}

\usepackage[svgnames]{xcolor}
\usepackage[pdftex]{hyperref,graphicx}
\usepackage{amsfonts,amssymb,amsmath}
\usepackage[english]{babel}
\usepackage[utf8]{inputenc}
\usepackage[T1]{fontenc}

\usepackage{xspace}

\usepackage{soul}

\usepackage{mathtools}
\usepackage{amsbsy}
\usepackage{amstext}

\usepackage[separate-uncertainty=false,per-mode=symbol]{siunitx}
\DeclareSIUnit\gauss{G}
\DeclareSIUnit\centimeter{cm}

\usepackage{braket}

\hypersetup{
    pdftitle = {Boson sampling with ultracold atoms in a programmable optical lattice},
    pdfauthor = {Carsten Robens, Iñigo Arrazola, Wolfgang Alt, Dieter Meschede, Lucas Lamata, Enrique Solano, and Andrea Alberti},
	colorlinks=true,linkcolor=blue,citecolor=blue,filecolor=blue,urlcolor=blue
  }

\newcommand{\figref}[2]{\hyperref[#1]{\ref{#1}(#2)}}

\addto\captionsenglish{}
\addto\captionsenglish{}

\renewcommand\textemdash{\leavevmode\unskip\kern0.8pt\rule[0.19\baselineskip]{8pt}{0.4pt}\kern1pt\ignorespaces}

\begin{document}

\title{Boson sampling with ultracold atoms in a programmable optical lattice}
\author{Carsten Robens}
\affiliation{Institut für Angewandte Physik, Universität Bonn, Wegelerstr.~8, 53115 Bonn, Germany}
\affiliation{MIT-Harvard Center for Ultracold Atoms, Research Laboratory of Electronics, and Department of Physics, Massachusetts Institute of Technology, Cambridge, Massachusetts 02139, USA}
\author{Iñigo Arrazola}
\affiliation{Department of Physical Chemistry, University of the Basque Country UPV/EHU, Apartado 644, 48080 Bilbao, Spain}
\affiliation{Vienna Center for Quantum Science and Technology, Atominstitut, TU Wien, 1040 Vienna, Austria}
\author{Wolfgang Alt}
\affiliation{Institut für Angewandte Physik, Universität Bonn, Wegelerstr.~8, 53115 Bonn, Germany}
\author{Dieter Meschede}
\affiliation{Institut für Angewandte Physik, Universität Bonn, Wegelerstr.~8, 53115 Bonn, Germany}
\author{Lucas Lamata}
\affiliation{Departamento de F\'isica At\'omica, Molecular y Nuclear, Universidad de Sevilla, 41080 Sevilla, Spain}
\affiliation{Instituto Carlos I de F\'isica Te\'orica y Computacional, 18071 Granada, Spain}
\author{Enrique Solano}
\affiliation{Kipu Quantum, Kurwenalstrasse 1, 80804 Munich, Germany}
\affiliation{IKERBASQUE, Basque Foundation for Science, Plaza Euskadi 5, 48009 Bilbao, Spain}
\affiliation{International Center of Quantum Artificial Intelligence for Science~and~Technology~(QuArtist) and Physics Department, Shanghai University, 200444 Shanghai, China}
\author{Andrea Alberti}
\affiliation{Institut für Angewandte Physik, Universität Bonn, Wegelerstr.~8, 53115 Bonn, Germany}
\affiliation{Fakultät für Physik, Ludwig-Maximilians-Universität München, 80799 München, Germany}
\affiliation{Max-Planck-Institut für Quantenoptik, Hans-Kopfermann-Straße 1, 85748 Garching, Germany}
\affiliation{Munich Center for Quantum Science and Technology, 80799 München, Germany}

\begin{abstract}
Sampling from a quantum distribution can be exponentially hard for classical computers and yet could be performed efficiently by a noisy intermediate-scale quantum device.
A prime example of a distribution that is hard to sample is given by the output states of a linear interferometer traversed by $N$ identical boson particles.
Here, we propose a scheme to implement such a boson sampling machine with ultracold atoms in a polarization-synthesized optical lattice.
We experimentally demonstrate the basic building block of such a machine by revealing the Hong-Ou-Mandel interference of two bosonic atoms in a four-mode interferometer.
To estimate the sampling rate for large $N$, we develop a theoretical model based on a master equation that accounts for particle losses, but not include technical errors.
Our results show that atomic samplers have the potential to achieve quantum advantage over today's best supercomputers  with $N \gtrsim 40$.
\end{abstract}

\maketitle

\section{Introduction}
\label{sec:intro}

The idea that quantum computers could efficiently solve problems that are believed to be intractable for classical computers is the main motivation for the development of quantum computing devices \cite{Blatt:2008,Ladd:2010,Weiss:2017}. However, despite the impressive progress in increasing the number of qubits and extending their coherence time \cite{Monz:2011,Kaufman:2015,Lekitsch:2017,Wang:2016,Barredo:2018a,Zeng:2017,Bernien:2017,Wang:2017,Bermudez:2017,Zajac:2017,Levine:2019,Chen:2021a,Pogorelov:2021}, building a fault-tolerant universal quantum computer is not yet within the reach of current quantum technology \cite{Wilczek:2016}.
This fact has stimulated the development of alternative concepts of quantum computation that can be performed with noisy intermediate-scale quantum (NISQ) devices \textemdash
machines that are far less demanding than a fault-tolerant quantum computer and yet can outperform the best available classical computers on specific tasks.
Examples of such NISQ devices are quantum annealers \cite{Das:2008}, quantum simulators \cite{Cirac:2012,Bloch-Dalibard:2012,Blatt:2012,Aspuru-Guzik:2012,Houck:2012,VanHoucke:2012,Georgescu:2014,Gross:2017,Ebadi:2021},
quantum learning machines \cite{Biamonte:2017,Benedetti:2019}, digital-analog quantum machines \cite{Parra-Rodriguez:2020}, and quantum sampling machines \cite{Lund:2017}.

A quantum sampling machine deals with the task of drawing from the probability distribution of the outcomes that are produced by measuring
a quantum system in a highly entangled state.
In essence, the idea is to use the randomness inherent to a measured quantum system to construct a hard-to-simulate sampling machine.
Compared to other problems (e.g., decision problems), quantum sampling has the advantage that its computational complexity can be ascertained for many quantum distributions \cite{Aaronson:2016a} by relying only on a few widely held assumptions (e.g., no collapse of the polynomial hierarchy).
Knowing the computational complexity of the specific problem allows us to gain important insights into the conditions (e.g., size of the Hilbert space) and class of quantum states \cite{GottesmanKnill:1999,Valiant:2002,Pednault:2017,Dalzell:2020} required to achieve a quantum advantage over classical machines \cite{Preskill:2013,Harrow:2017}.
Quantum sampling also appeals for practical reasons because its computational hardness is generally  robust to small experimental errors \cite{Bouland:2018,Dalzell:2021}. 
Such a natural tolerance to errors makes quantum sampling a particularly suitable task to be performed with NISQ devices.
Based on these motivations, several proposals have been put forward, where one draws samples from the state generated by a quantum circuit such as: constant-depth quantum circuits \cite{Terhal:2004,Bermejo-Vega:2018,Bravyi:2018}, instantaneous quantum polynomial-time circuits \cite{Shepherd:2009,Bremner:2016}, random quantum circuits \cite{Boixo:2018,Neill:2018,Bouland:2018,Arute:2019,Wu:2021}, and linear quantum circuits of indistinguishable bosons \cite{Aaronson:2011}, better known as boson samplers.

Boson sampling \cite{Gard:2015}
refers to the problem of sampling from the probability distribution of the outcomes generated by $N$ identical, noninteracting bosons that have travelled through an $M$-mode interferometer, with the initial and final $N$-particle states being of the form of Fock states.
The state of each particle evolves according to the same $M{\times}M$ unitary matrix $U$, mapping the initial into the final $M$ modes of the linear interferometer.
The probability of detecting a particular outcome comprising $N$ bosons is proportional to the absolute square of the permanent of a $N{\times}N$ submatrix of $U$ \cite{Scheel:2004,Lim:2005}.
In spite of its compact analytical expression, the permanent (and likewise its absolute square) is very hard to compute, for it requires a time exponential in $N$ \cite{Glynn:2010}.
In fact, even its approximation to a multiplicative factor has been shown \cite{Aaronson:2011b} to fall into the \#P-hard complexity class \cite{Valiant:1979}.
From a physics point of view, it is worth emphasizing that the hardness of this problem is due to the quantum statistics of indistinguishable bosons and not to interactions between particles \cite{Aaronson:2013,Gard:2014}.

A number of experiments have been reported demonstrating boson sampling in photonic quantum circuits \cite{Broome:2013,Tillmann:2013,Spring:2013,Crespi:2013,Carolan:2014,Spagnolo:2014,Tillmann:2015,Carolan:2015,Loredo:2017,Wang-Hui:2017,Wang:2018,Brod:2019}, with the current record being $N\,{=}\,20$ photons in $M\,{=}\,60$ modes \cite{Wang:2019e}.
Because of losses, however, it is hard to reach in the near future a much higher number of photons in a deterministic manner.
This limitation along with the development~\cite{Neville:2017} of more efficient classical algorithms for simulating boson sampling
have prompted the study of variants of the problem that better cope with losses, such as lossy boson sampling \cite{Aaronson:2016b,Paesani:2019}, scattershot boson sampling \cite{Lund:2014,Bentivegna:2015,Zhong:2018,Paesani:2019} and Gaussian boson sampling \cite{Hamilton:2017,Paesani:2019}.
This latter in particular, which uses squeezed light instead of single photons, has demonstrated \cite{Zhong:2020,Zhong:2021,Madsen:2022,Deng:2023} a huge increase in the number of photons detected at the interferometer output, on the order of 100, leading to claims of a quantum advantage.

The quantum advantage of Gaussian boson sampling machines has been questioned, though, as it has been shown that classical sampling algorithms are able to efficiently draw samples from a sufficiently close distribution \cite{Popova:2021,Bulmer:2021,Villalonga:2021,Oh:2023}.
For random quantum circuits \cite{Arute:2019,Wu:2021}, likewise, effective representations of the qubits' entangled state have been found \cite{Huang:2020a,Zhou:2020a} using tensor networks, which result in a tremendous speed-up of classical simulations, since only a tiny fraction of the Hilbert space is actually used when the gate fidelity is below a certain threshold.
Such a race between quantum hardware and ever more efficient classical algorithms is indeed expected to continue in the coming years, promising new insights into what makes quantum systems advantageous from a computational complexity perspective.
Remarkably, it was shown \cite{Dalzell:2020} based on fine-grained complexity arguments that a boson sampling quantum machine of the original type \cite{Aaronson:2011} with $N\,{=}\,100$ and $M\,{=}\,500$ achieves quantum advantage with respect to any (i.e., known and unknown) classical simulation algorithms.
These numbers are large, yet not beyond the reach of scalable NISQ devices such as ultracold atoms in optical lattices.

In this article, we propose to use ultracold atoms in state-dependent optical lattices as a scalable architecture for boson sampling with hundreds of bosons.
We also report on the experimental realization of the basic building block of the proposed boson-sampling machine, demonstrating the Hong-Ou-Mandel interference between two atoms trapped in state-dependent optical lattices.
In our scheme, atoms cooled into their motional ground state play the role of identical bosons, while the lattice site as well as two internal atomic states serve as the bosonic modes.
Distant modes, associated with different lattice sites, are brought together by state-dependent shift operations, which are realized with polarization-synthesized optical lattices \cite{Robens:2016pol,Robens:2017}.
Modes that are spatially overlapped are coupled in pairs, by a combination of microwave and site-resolved optical pulses, realizing the analog of phase-programmable photonic quantum circuits \cite{Silverstone:2016,Russell:2017}.

It should be mentioned that based on a similar motivation to establish a quantum advantage, other NISQ proposals alternative to photonic boson samplers have been put forward in the past years relying on trapped ions~\cite{Shen:2014,Chen:2023,Katz:2023}, superconducting circuits~\cite{Peropadre:2016}, and neutral atoms with microwave assisted tunneling~\cite{Muraleedharan:2019}.
Furthermore, after the submission of this article, the first demonstration of boson sampling using ultracold atoms has been presented~\cite{Young:2023}, involving 180 atoms spread across 1000  sites in a tunnel-coupled optical lattice.
This recent experiment beautifully demonstrates the potential of ultracold atoms to realize a large-scale boson sampling device, once arbitrarily programmable quantum circuits can be created.
Such arbitrary quantum circuits could be produced either using programmable tweezer arrays~\cite{Young:2022} or with state-dependent potentials, as discussed in this article.

\begin{figure*}[tbp]
	\includegraphics[width=\textwidth]{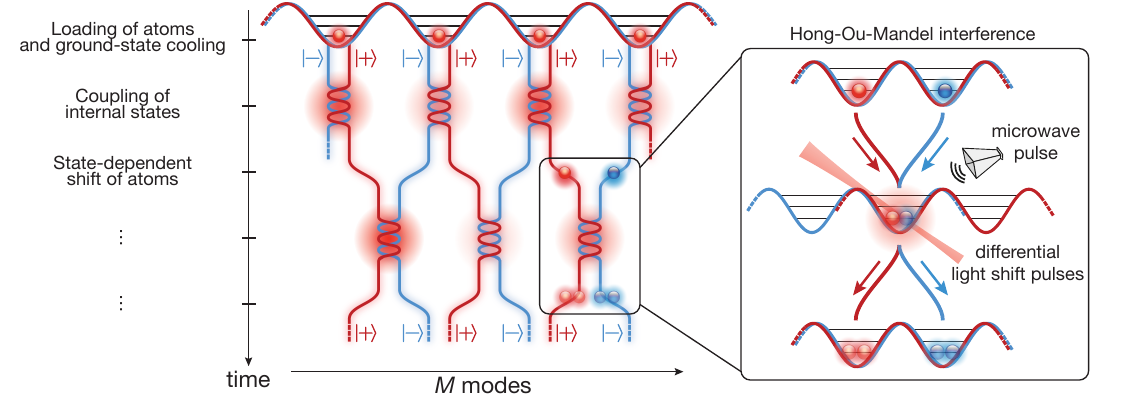}
	\caption{\label{fig1}
	 \textbf{Illustration of a quantum circuit based on ultracold atoms in state-dependent optical lattices.}
	 Each site accommodates two modes, $\ket{\pm}$, of the quantum circuit represented by two internal states of the atom.
	 A representative initial state is shown, where every second mode of the circuit is occupied.
	 State-dependent shift operations connect distant modes, while local operations couple the internal modes.
	 	 Inset: Hong-Ou-Mandel interference of two indistinguishable atoms, where a combination of local phase-imprinting pulses and microwave pulses realize the equivalent of a generalized photonic beam splitter.
	 	 }
\end{figure*}

\section{Boson sampling with atoms in optical lattices}\label{PSOLforQC}

A boson sampling quantum machine is in essence an $M$-port quantum circuit traversed simultaneously by $N$ identical bosons that do not interact with each other.
As there are no interactions between the particles, such a quantum circuit behaves as a linear interferometer, mapping each input mode into a superposition of the output modes,
\begin{equation}
    a_i^\dagger \rightarrow \sum_{j=1}^{M} {U}_{ji} \,a_j^\dagger\,.
\end{equation}
Here, $a^\dagger_i$ is the operator creating a boson in the \mbox{$i$-th} mode, and $U_{ji}$ is the matrix element of a unitary transformation $U$, which is
randomly chosen from the uniform distribution (i.e., Haar measure) over all $M\,{\times}\,M$ unitaries.
The randomness of $U$ ensures that no particular feature can be exploited to efficiently simulate the boson sampler machine with a classical computer.

By detecting the occupation of the output modes, the machine thus directly samples from the probability distribution $P(n_1,n_2,\ldots, n_M)=\left| \langle n_1,n_2,\ldots,n_M \right| U \left| \psi_0 \rangle \right|^2$. 
Here, $n_i$ denotes the number of bosons in the $i$-th output mode, and $\ket{\psi_0}$ represents the initial state with $N$ identical bosons, each occupying a particular input mode.
According to best-known algorithms \cite{Neville:2017}, sampling from $P$ cannot be performed efficiently with classical computers, as it is bound to computing the permanent of $N\times N $ matrices, which requires a computation time of order $\mathcal{O}(N^2 2^N)$ \cite{Wu:2018}.

Importantly, to be hard to simulate by a classical computer, a boson sampler must have a number of modes much larger than that of particles, $M\gg N \gg 1$ \cite{Aaronson:2011}.
The gold standard satisfying this condition is given by the scaling law $M=N^2$
because it ensures that detecting two or more particles in any of the output modes has a small probability \cite{Arkhipov:2012}.
In fact, only when the output modes are singly occupied, i.e., for the so-called collision-free outcomes, does the conjectured hardness of boson sampling hold \cite{Aaronson:2011}.
We therefore assume such a quadratic scaling in this paper.
It should, however, be emphasized that this scaling has so far only been experimentally realized with a relatively small number of particles, $N<10$, by photonics devices \cite{Wang:2019e}.

In the remainder of this section, we develop a concept how to implement such a boson sampling quantum machine with ultracold atoms in state-dependent optical lattices.
We start with the key idea of how to construct arbitrary quantum circuits, and then discuss initialization and detection.

\subsection{Arbitrary quantum circuits using polarization-synthesized optical lattices}
\label{sec:wiring_qnt_circuits}

Figure~\ref{fig1} illustrates how to ``wire'' an arbitrary quantum circuit using ultracold atoms in state-dependent optical lattices.
The idea here is to use the lattice sites along with two internal states of the atom, $\ket{+}$ and $\ket{-}$, to represent the modes of the quantum circuit,
so that $M/2$ lattice sites accommodate $M$ modes.
Modes associated with adjacent sites are connected in pairs by state-dependent shift operations.

Such state-dependent shift operations  can be performed using polarization-synthesized optical lattices \cite{Robens:2016pol}.
This piece of technology relies on the synthesis of polarization states of light to create two movable, fully independent periodic potentials,
\begin{equation}
V_\pm(x) = V^0_\pm\cos^2\{2\pi/\lambda_L[x-x_\pm(t)]\},
\end{equation}
which selectively trap atoms in either the $\ket{+}$ or $\ket{-}$ internal state.
Here, $V^0_\pm$ represents the trap depth and $x_\pm(t)$ the position of the respective periodic potentials, $x$ is the coordinate along the lattice direction, and $\lambda_L$ the wavelength of the lattice laser.
The underlying concept of state-dependent optical potentials is suited to atomic species such as Rb and Cs \cite{Deutsch:1998,Jaksch:1999}.
In this work, we will consider specifically the case of $^{133}$Cs, where the internal modes are the hyperfine states $\ket{+}=\ket{F=4,m_F=3}$ and $\ket{-}=\ket{F=3,m_F=3}$, and $\lambda_L$ is set to the value of $\SI{870}{\nano\meter}$, for which the trapping potential of right and left circularly polarized light selectively trap the two internal states.

The lattice potentials must be chosen sufficiently deep, with $V^0_\pm$ being of order of a few hundred recoil energies, to prevent atoms from tunneling to the neighboring sites.
In such a deep-lattice regime, one can shift the atoms to the adjacent sites in a state dependent manner by simply varying the relative position, $x_+(t)\,{-}\,x_-(t)$, as a function of time $t$.
We have experimentally demonstrated \cite{Robens:2015} that repeated state-dependent shift operations preserve the coherence between the two internal states.
Furthermore, we have shown in a recent work \cite{Lam:2021} that shifting the atoms by one lattice site can be rapid, with the minimum duration being bounded by the trap period at around $\SI{10}{\micro\second}$.

Crucially, a quantum circuit such as the one in Fig.~\ref{fig1}, where the modes are locally coupled in alternating pairs, allow one to realize any arbitrary $M{\times}M$ unitary transformation $U$ of the input into the output modes \cite{Reck:1994,Clements:2016}.
For a generic matrix $U$, a minimum number $M(M{-}1)/2$ of local operations ${T}$ is required, arranged in a circuit of $M$-step depth \cite{Clements:2016}.
Such an operation $T$ defines the basic unit of the quantum circuit, coupling together the modes $\ket{\pm}$ associated with a given lattice site,
\begin{equation}
    \label{eq:basiccoupling}
    T=\left(\begin{array}{cc} e^{-i\phi}\cos(\theta/2)&-\sin(\theta/2)\\e^{-i\phi}\sin(\theta/2)&\cos(\theta/2) \end{array}\right),
\end{equation}
where $\phi$ and $\theta$ are parameters depending on the particular site and time step.
A programmable quantum device of this kind is said to be completely controllable \cite{Schirmer:2001}.

We propose to implement $T$ through composite pulses, where two types of elementary operations are stacked together: local phase imprints $A(\varphi)$ and global Hadamard pulses $H$.
In fact, for any $T$, one can find suitable angles $\theta$ and $\phi$ yielding the following decomposition:
\begin{equation}
    \label{eq:decomposition}
T = e^{-i\phi/2}\,T'= e^{-i\phi/2} H^\dagger A(\theta) H A(\phi),
\end{equation}
where ${A}(\varphi)=\exp[-i {\sigma}_z \varphi/2]$ imprints onto the two modes $\ket{\pm}$ a relative phase depending on the lattice site, whereas $H = \exp(-i {\sigma}_x \pi/4)$ acts on all sites identically, realizing the equivalent of a beam splitter (here, ${\sigma}_i$ represent the Pauli matrices).
Note that the local common-mode phase shift by $\phi/2$, which appears in Eq.~(\ref{eq:decomposition}), can be avoided by conveniently adapting the algorithm by Clements \emph{et al.}~\cite{Clements:2016} to use $T'$ instead of $T$ as the basic unit.

The decomposition of ${T}$ in Eq.~(\ref{eq:decomposition}) reveals a direct analogy to phase-programmable photonic circuits \cite{Silverstone:2016,Russell:2017}.
Their structure reveals however an important difference: for ultracold atoms, a single spatial dimension suffices to wire the circuital modes, whereas at least two dimensions are necessary for photonic devices \cite{Brod:2019}.
The advantage of ultracold atoms simply arises from the fact that massive particles can be held in a specific location by a trapping potential.

The global Hadamard gates are readily implemented by means of $\pi/2$ microwave pulses, which act homogeneously on all lattice sites and require a time of order of \SI{1}{\micro\second}.
The local operator ${A}(\varphi)$ can be realized by exploiting the differential Stark shift that is produced by an array of laser beams focused on the target lattice sites through a high-numerical-aperture objective lens \cite{Weitenberg11,Preiss:2015,Robens:2016mbg}.
Exploiting the vector polarizability of alkali atoms \cite{Deutsch:1998}, one can imprint a purely differential phase shift $\varphi$ onto the atoms by means of a circularly polarized light field.
For Cs atoms, this condition is fulfilled when the wavelength of the addressing light field is chosen at $\lambda_A=\SI{880}{\nano\meter}$.
Such local pulses also require that the addressing beam has a nonzero component along the quantization axis.
This additional condition can be met by tilting the quantization axis with respect to the lattice direction (see Appendix~\ref{app:q_axis_tilt}).

The differential phase shift $\varphi$ is directly controlled by the product of the laser intensity and pulse duration.
We estimate that the addressing pulses ${A}(\varphi)$ can be realized in about $\SI{1}{\micro\second}$ using approximately $\SI{10}{\micro\watt}$ of laser power per addressed lattice site.
These pulses have a small impact on the coherence time of the atoms, because the probability that an atom scatters a photon off the addressing beam is approximately $\num{4e-5}$.

It is worth noting that the composite pulse scheme proposed to implement $T$ offers an important advantage over other schemes that use local resonant pulses to couple the two hyperfine states, $\ket{\pm}$.
The reason for this is the difference in sensitivity to the crosstalk caused by the light field at the sites adjacent to the target lattice site.
Differential Stark shift pulses, as in the proposed scheme, depend on the intensity of the addressing laser beams in their Hamiltonian, while resonant pulses directly depend on the respective electric fields.
This different sensitivity implies that  for the same intensity $I$ leaking to the neighboring sites, crosstalk errors are smaller in the proposed scheme:
the crosstalk infidelity is proportional to $I^2$ in the proposed scheme, in contrast to $I$ in the resonant pulse schemes.

\subsection{Initializing an array of identical atoms}

Atom sorting techniques have been demonstrated~\cite{Barredo:2016,Endres:2016,Kim:2016,Robens:2017,Kumar:2018,Ohl-de-Mello:2019} where movable optical potentials are used to deterministically fill a predefined array of optical traps with one atom each.
Recently, it has become possible to achieve densely-packed arrays with more than 1000 atoms~\cite{Gyger:2024,Norcia:2024}
Once loaded into the desired sites of an optical lattice, atoms can be cooled to their motional ground state using sideband cooling techniques, making them indistinguishable in their mechanical degrees of freedom.
Ground state probabilities above $\SI{90}{\percent}$ have already been achieved for atoms trapped in optical tweezers \cite{Kaufman:2012,Kumar:2018,Yu:2018,Jenkins:2022}, whereas higher values above $\SI{99}{\percent}$ are expected \cite{Li:2012} for more tightly confined atoms in a three-dimensional optical lattice.

\subsection{Detection of individual atoms}

The final state is measured by recording a fluorescence image of the atoms \cite{Alberti2015,Robens:2016mbg}.
Using a high-resolution objective lens, the positions of the individual atoms in the optical lattices can be reconstructed with a high fidelity, exceeding \SI{99}{\percent}.
Standard fluorescence-imaging techniques, however, only give information about the occupation of modes belonging to distinct lattice sites.
To also resolve the occupation of modes associated with the same site, a state-sensitive detection scheme resolving the two hyperfine states, $\ket{\pm}$, is needed.
For this purpose, a long-distance state-dependent transport operation can be used to realize an optical Stern-Gerlach detection, mapping the two internal states to different lattice sites \cite{Robens:2016b,Wu:2019a}.
Alternatively, one can use a magnetic Stern-Gerlach detection scheme in a multilayer optical lattice \cite{Preiss:2015b}.

A particle number resolving detection is more demanding.
Standard fluorescence imaging is not suitable because of light-induced collisions, which only allow the parity of the occupation number to be measured \cite{Bakr:2009}.
One approach is to distribute the atoms to multiple sites prior to fluorescence imaging \cite{Rispoli:2019}, similar to spatial multiplexing in photonic devices \cite{Carolan:2014}.
An improved version of this approach consists in using a pinning lattice to detect the atoms \cite{Omran:2015,Lukin:2018}.
Alternatively, one can exploit interaction blockade to induce occupation-dependent tunneling to distinct sites of an optical lattice \cite{Preiss:2015b}.

\section{Scaling law of an atom-boson-sampling machine}\label{ScalingAndErrors}

To appreciate the quantum advantage of an atom boson sampler over classical machines, we study in this section  the sampling rate, $R$, as a function of the number of bosons.

\subsection{Ideal boson sampler}

In the ideal case of no atom losses, the sampling rate is simply given by the repetition rate $R_\text{rep}$ at which $N$ bosons are made to interfere with each other in a quantum circuit of $M$ modes.
Three terms determines this rate, 
\begin{equation}
    \label{eq:samp_rate_ideal}
    R_\text{rep}^{-1} =  t_\text{init} + t_\text{exec} + t_\text{det},
\end{equation}
where $t_\text{init}$ is the time for preparing the initial state of $N$ atoms, $t_\text{exec}$ is the time for executing the quantum circuit, and $t_\text{det}$ is the time for detecting the atoms in the $M$ output modes.

To estimate the execution time $t_\text{exec}$, we assume that local Stark shift pulses can be carried out in parallel by optically addressing all sites simultaneously \cite{Pogorelov:2021,Christen:2022a}.
The time for executing the quantum circuit is thus proportional to the total number of steps, which in turn is equal to the number of modes $M$ (see Sec.~\ref{sec:wiring_qnt_circuits}).
Considering that the number of modes is given by $M=N^2$, as previously reasoned, we find that the execution time has a quadratic dependence on the number of bosons, $t_\text{exec} = N^2 t_\text{step}$, where $t_\text{step}$ is the time to perform the single step.

The initialization time $t_\text{init}$ is proportional to $N$ for linear atom sorting \cite{Endres:2016,Barredo:2016}, and to $\log(N)$ for parallel atom sorting \cite{Robens:2017}.
For simplicity, we assume that this time is fixed at $\SI{500}{\milli\second}$, since the initialization of $100$ atoms can be efficiently performed in less than this time \cite{Ebadi:2021}.
Likewise, we consider the detection time to be fixed, $t_\text{det}=\SI{100}{\milli\second}$, since both positions and spin can be efficiently detected for all atoms in a single operation relying on fluorescence imaging \cite{Robens:2016b,Wu:2019a}.

Furthermore, if we post-select only those events with all atoms populating a different output mode (i.e., the so-called collision-free events), because these are the events hard to simulate with a classical machine, the sampling rate is reduced by a constant factor $1/e$ in the limiting case of large $N$ \cite{Arkhipov:2012}, leading to $R_\text{ideal} \approx R_\text{rep}/e$.

Thus, we conclude that under ideal conditions, an atom boson sampler can draw events from the boson distribution efficiently, since its computation time $1/R_\text{ideal}$ scales with $N^2$ for sufficiently large $N$ (i.e., polynomial time complexity).
In contrast, classical computer simulations require an exponentially longer time to perform the same task, which scales with $\mathcal{O}(N^2 2^N)$ \cite{Neville:2017} (see Sec.~\ref{PSOLforQC}).

\def\SymbolFigScaling#1#2#3{\raisebox{#1}[0pt][0pt]{\includegraphics[page=#3,height=#2\baselineskip]{Fig2_symbols.pdf}}}

\newsavebox{\redcircle}
\savebox{\redcircle}{\SymbolFigScaling{0pt}{0.6}{1}}

\newsavebox{\bluesquare}
\savebox{\bluesquare}{\SymbolFigScaling{-0.2pt}{0.5}{2}}

\newsavebox{\emptydiamond}
\savebox{\emptydiamond}{\SymbolFigScaling{-0.2pt}{0.55}{5}}

\newsavebox{\emptystar}
\savebox{\emptystar}{\SymbolFigScaling{0pt}{0.6}{4}}

\newsavebox{\emptyhexagram}
\savebox{\emptyhexagram}{\SymbolFigScaling{-0.2pt}{0.6}{3}}

\newsavebox{\emptytriangle}
\savebox{\emptytriangle}{\SymbolFigScaling{0pt}{0.5}{7}}

\begin{figure}[tbp]
    \includegraphics[width=1\columnwidth]{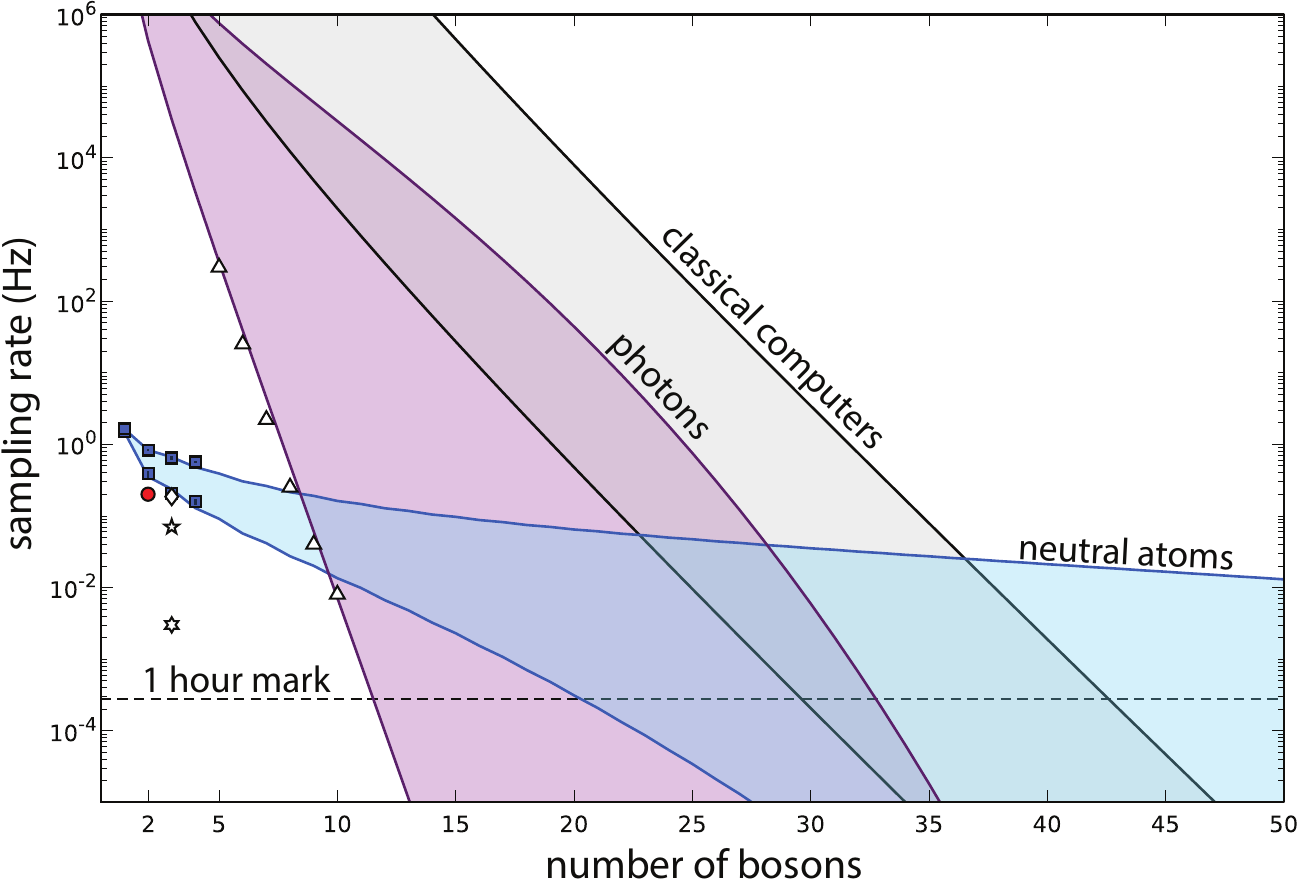}
    \caption{\label{fig2}
\textbf{Scaling of boson sampling machines.} Predicted sampling rate versus the number of bosons $N$ is shown for atomic (blue),
photonic (purple), and classical (grey) boson sampling machines.
The atomic sampling is computed based on Eq.~(\ref{AtomicScaling}),
where an extended model (see appendix~\ref{app:Pair}) is used for $P_{\rm step}$ to account for both particle pairs and triplets. Our atomic sampling model disregards decoherence by technical noise sources; see text.
The photonic and classical curves correspond to Eqs.~(\ref{photonscaling})~and~(\ref{computerrate}), respectively.
Under state-of-the-art conditions, atomic NISQ devices have the potential to overtake classical algorithms at $N\approx 37$.
Pioneering photonic experiments are marked with symbols \usebox{\emptydiamond}~\cite{Broome:2013},
\usebox{\emptystar}~\cite{Tillmann:2013}, and \usebox{\emptyhexagram}~\cite{Spring:2013},
while the best reported boson sampling experiment with \usebox{\emptytriangle}~\cite{Wang:2019e}.
Note that the photonic experimental points refer to implementations of the original boson sampling problem, and thus do not include Gaussian boson sampling experiments, for the reasons presented in Sec.~\ref{sec:intro}.
Our proof-of-principle experiment, described in Sec.~\ref{HOMExpSect}, is marked with \usebox{\redcircle} symbol.
The numerical results obtained with the master equation approach (see Appendix~\ref{App:ExactSim}) are indicated with \usebox{\bluesquare} symbols, with the vertical bars indicating the 1-$\sigma$ statistical uncertainty.
}
\end{figure}

\subsection{NISQ boson sampler}

In a realistic scenario typical of NISQ devices, the sampling rate is significantly degraded by state preparation errors, atom losses while executing the quantum circuit, and detection inefficiency.
Along the lines of Refs.~\cite{Wang-Hui:2017,Neville:2017}, we estimate the sampling rate as:
\begin{equation}\label{AtomicScaling}
R_\text{NISQ} = (\eta_\text{init} \eta_\text{det})^{N} P_\text{surv} \, R_\text{ideal},
\end{equation}
where $\eta_\text{det}$ is the detection efficiency, $\eta_\text{init}$ the cooling efficiency into the motional ground state, and $P_\text{surv}$ is the probability that all $N$ atoms survive.
As reasoned in Sec.~\ref{sec:intro}, we only consider the case where no particle is lost. 

Equation~(\ref{AtomicScaling}) immediately reveals that the computation time $1/R_\text{NISQ}$ scales exponentially with the number of particles.
It is therefore important to carefully evaluate the expression in Eq.~(\ref{AtomicScaling}) to determine the conditions when an atom boson sampler offers a quantum advantage over classical computer simulations.

As described in Sec.~\ref{PSOLforQC}, cooling and detection of ultracold atoms can be done efficiently, with reported values of $\eta_\text{init}$ and $\eta_\text{det}$ above \SI{90}{\percent}~\cite{Kaufman:2012,Kumar:2018,Yu:2018,Jenkins:2022} and \SI{99}{\percent}~\cite{Robens:2016b}, respectively.
The survival probability $P_\text{surv}$ in Eq.~(\ref{AtomicScaling}) depends on two main loss mechanisms, which we discuss below.

The first mechanism responsible for the loss of atoms is collisions between one of the $N$ trapped atoms and a particle from the background gas at room temperature, causing the atom involved in the collision to be ejected from the trap.
For $N$ atoms, the probability that no collision with the background gas occurs during a single step is given by the exponential formula $P_\text{step}^\text{BG} = \exp{(-N t_{\text{step}}/\tau_{\rm BG})}$, where $\tau_{\rm BG}$ is the background-gas-limited mean lifetime of a single atom.

The second mechanism leading to the loss of atoms is given by inelastic collisions of atoms occupying the same lattice site.
Inelastic collisions cause the atoms to change their hyperfine state \cite{Mies:1996,Leo:1998,Chin:2000}, and to acquire kinetic energy, thus leaving the motional ground state where they are initially prepared.
Such inelastic collisions typically result in the loss of atoms from the trap because of the large energy separation between hyperfine states (several $\si{\mega\hertz}$ between adjacent $m_F$ states, $9.2\,\si{\giga\hertz}$ between different $F$ states, expressed in frequency units).
Three-body collisions are neglected because the probability of three (and more) particles occupying the same lattice site compared to that of pairs is negligible in the limit of a large number of modes (see Appendix~\ref{app:Pair}).
To account for the two-atom lossy collisions, we introduce the survival probability of a pair of atoms located in the same lattice site, given by $\exp{(-t/\tau_{\rm TB})}$, where $\tau_{\rm TB}$ is the mean lifetime limited by two-body collisions.
Note that for simplicity we use the same constant $\tau_\text{TB}$ without differentiating between the three possible spin configurations of the two bosons occupying the same site.  
To estimate the number of pairs of atoms that can collide onsite, we make the conservative assumption that all states of $N$ bosons have equal probability of being occupied at every time step.
This is a conservative assumption, which overestimates the probability of inelastic collisions at the initial steps since the atoms are first prepared in different sites.
Based on these assumptions,
the probability of finding $k$ sites occupied by exactly a pair of atoms can be estimated as $P_k=e^{-3/2}(3/2)^k/k!$ (see Appendix~\ref{app:Pair}).
For example, $P_0$ is the probability of the collisionless subspace, where all atoms occupy distinct lattice sites.
Thus, the survival probability per time step, limited by two-body lossy collisions, is obtained by the evaluating the following sum,
\begin{equation}\label{SurvProb1}
P_{\rm step}^\text{TB} = \sum_{k=0}^{N/2} P_ke^{-k\frac{t_{\rm step}}{\tau_{\rm TB}}} \approx \exp{\left[\frac{3}{2}\left(e^{-\frac{t_{\rm step}}{\tau_{\rm TB}}}-1\right)\right]},
\end{equation}
where the expression on the right-hand side holds in the limit of large $N$ (see Appendix~\ref{app:Pair}).
For weak two-body losses, $t\ll \tau_\text{TB}$, the survival probability $P_{\rm step}^\text{TB}$ decays as $\exp{[-3t_{\rm step}/(2\tau_{\rm TB})]}$, while in the limiting case of strong losses the survival probability approaches $P_0$.

Combining the two loss mechanisms, the survival probability per step is simply given by  $P_\text{step} = P_{\rm step}^\text{BG} P_{\rm step}^\text{TB}$.
Because the execution of the quantum circuit requires $M=N^2$ steps, the total survival probability of $N$ atoms (where $N\gg1$) is thus
\begin{align}\label{SurvProb2}
P_{\rm surv}&=(P_{\rm step})^M  \\
&= \exp\{-N^3 t_{\text{step}}/\tau_{\rm BG}+N^2 (3/2)(e^{-\frac{t_{\rm step}}{\tau_{\rm TB}}}-1) \}. \nonumber
\end{align}
A comparison of the two terms in Eq.~(\ref{SurvProb2}) shows that under typical conditions $t_\text{step}\ll\tau_\text{TB}\ll \tau_\text{BG}$  two-body collisions are the dominant loss mechanism for $N < N_\text{threshold} = 3\tau_\text{BG}/(2\tau_\text{TB})$.
As will be argued below, realistic experiments are expected to operate with a number of atoms below this threshold.

To evaluate $R_\text{NISQ}$ in Eq.~(\ref{AtomicScaling}), we consider two different scenarios, which are based on conservative and state-of-the-art assumptions, respectively.
In the conservative scenario, we assume a step duration $t_{\rm step}=\SI{170}{\micro\second}$ and a mean lifetime limited by two-body collisions $\tau_{\rm TB}=\SI{40}{\milli\second}$, while in the state-of-the-art scenario, we consider $t_{\rm step}=\SI{33}{\micro\second}$ and $\tau_{\rm TB}=\SI{400}{\milli\second}$.
For the initialization and detection times, we assume $t_{\rm init}=\SI{500}{\milli\second}$ and $t_{\rm det}=\SI{100}{\milli\second}$, with efficiencies of $\eta_{\rm init}=0.90$ and $\eta_{\rm det}=0.99$ for the conservative scenario, and $\eta_{\rm init}=\eta_{\rm det}=0.99$ for the state-of-the-art scenario.
For the mean lifetime limited by background gas collisions, we take $\tau_{\rm BG}=\SI{360}{\second}$ in both scenarios.
The threshold value $N_\text{threshold}$ is larger than $1000$ atoms in both scenarios, implying that for a realistic number of atoms the dominant loss mechanism is inelastic two-body collisions rather than collisions with the background gas.

In Fig.~\ref{fig2}, we show the sampling rate $R_\text{NISQ}$ as a function of the number of particles $N$, computed for an atom boson sampler with  conservative and state-of-the-art assumptions (blue curves).
To identify the regime of potential quantum advantage, we present in the same figure the sampling rate $R_\text{classical}$ of best algorithms \cite{Neville:2017} simulating a boson sampler using a standard laptop and the Tianhe-2 supercomputer (gray curves).
For the sake of comparison, we also report in the figure the sampling rate expected for NISQ photonic devices (purple curves) for a conservative and state-of-the-art scenario; see Appendix~\ref{app:OtherPlat} details.

To validate our model of the sampling rate $R_\text{NISQ}$ in Eq.~(\ref{AtomicScaling}), we have carried out exact numerical simulations based on a master equation approach (Appendix~\ref{App:ExactSim}) for $N$ up to 4.
The result of the numerical simulations (blue squares in Fig.~\ref{fig2}) shows a very good agreement with the curves from the model.

In the scenario of $N=50$, which is relevant for quantum advantage, we find that the sampling rate of the fastest atomic sampler proposed in this work ($t_\text{step}=\SI{33}{\micro\second}$) is dominated by initialization and detection times.
We expect a different situation in tunnel-coupled optical lattices~\cite{Young:2023}, where a tunneling event (i.e., the equivalent of a time step) takes about $t_\text{step}\approx \SI{1}{\milli\second}$.
For these samplers, the execution time $t_\text{exec}=N^2\,t_\text{step}$ is of order of a few seconds for $N=50$ particles and will likely be the factor limiting the sampling rate.

We should emphasize that our model takes into account interactions between atoms only through their dissipative effects, which are accounted for in Eq.~(\ref{AtomicScaling}) in terms of atom losses.
The model does not consider interaction-induced phase shifts.
Modeling coherent interactions between atoms requires a different approach \cite{Brunner:2018}, and goes beyond the scope of this work.
We expect that the interaction-induced phase shifts will make this boson sampling problem even harder to simulate on classical computers compared to its linear counterpart, as was originally proposed \cite{Aaronson:2011b}.

Finally, we would like to point out that, both in our theoretical model of Eq.~(\ref{AtomicScaling}) and in the numerical simulations, we have not considered technical errors affecting the operations during the quantum circuit.
Future investigations on the scaling of atomic sampling devices will have to address not only atom losses, but also the loss of coherence.
The specific technical errors will depend on the specific implementation, and a detailed analysis thereof goes beyond the scope of this work. We refer the reader to the appendix of Ref.~\cite{Alberti14} for a discussion of the main technical decoherence mechanisms, and to \cite{Robens:2016pol} for a characterization of the noise sources affecting polarization-synthesized optical lattices.
The present work rather focuses on quantifying the effect of fundamental errors that exist across all neutral atom platforms.

\section{Experimental demonstration of the Hong-Ou-Mandel interference}
\label{HOMExpSect}

We have performed a proof-of-principle experiment with two atoms in a four-mode interferometer, which demonstrates the Hong-Ou-Mandel effect with atoms, as schematically shown in the inset of Fig.~\ref{fig1}.
Such an experiment establishes the basic building block of the envisaged boson sampling machine.

The Hong-Ou-Mandel effect with atoms has been previously demonstrated experimentally using movable optical tweezers \cite{Kaufman14}, an optical lattice superimposed to a box potential \cite{Islam:2015}, and a free-fall atom interferometer \cite{Lopes:2015cna}.
Compared to these setups \cite{Kaufman:2018} and to related proposals based on microwave-induced tunneling \cite{Muraleedharan:2019}, our setup is distinguished by the way modes are coupled, where the atoms are moved with state-dependent shift operations \cite{Robens:2016pol} instead of having them tunnel through an optical potential barrier.
Our approach enables faster operations on the scale of few microseconds instead of milliseconds.

\begin{figure}[b]
    \includegraphics[width=1\columnwidth]{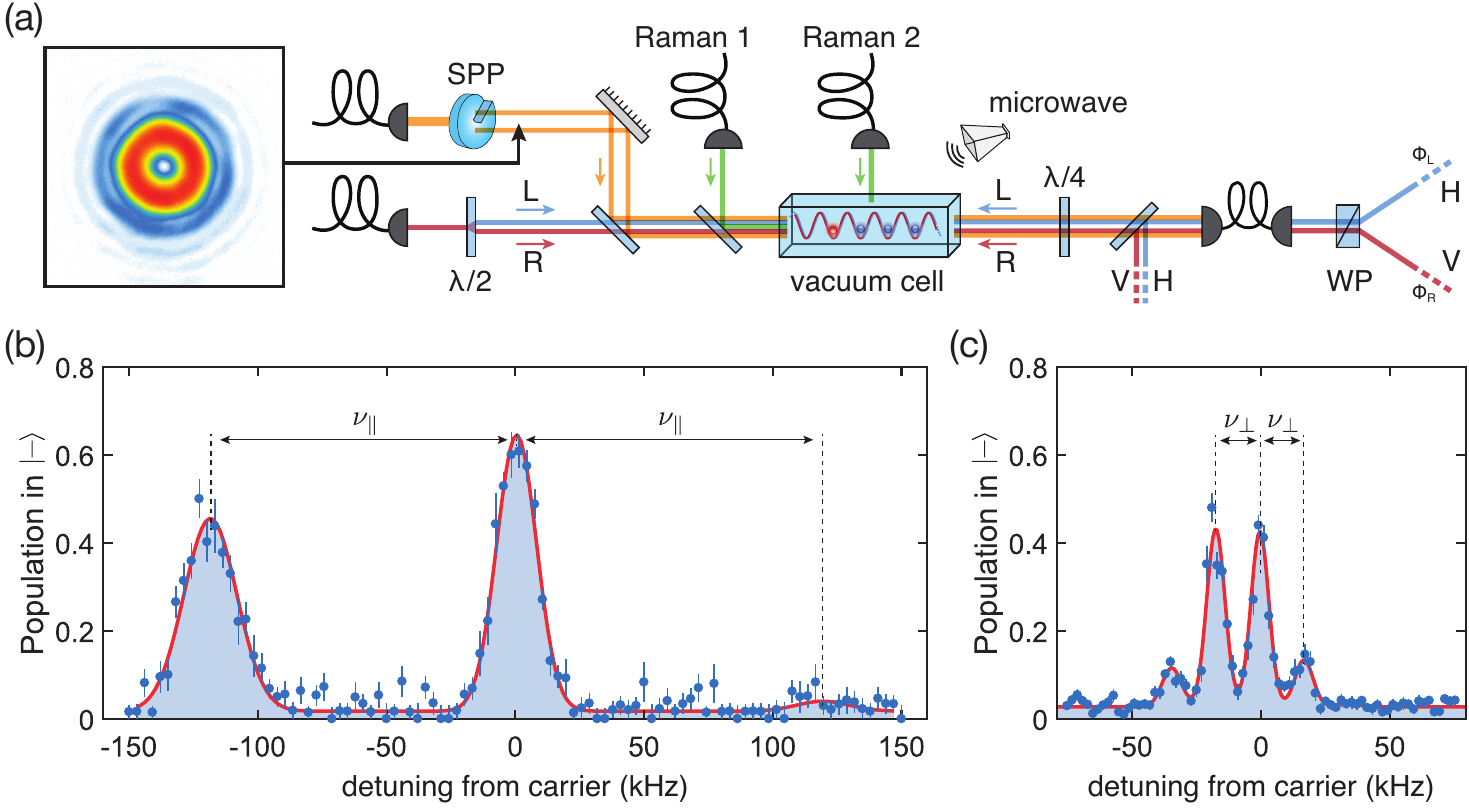}
    \caption{\label{fig4}
    \textbf{Experimental setup probing the Hong-Ou-Mandel interference with two atoms.} (a) Atoms are trapped in a polarization-synthesized optical lattice, formed by two optical standing waves of left (L) and right (R) circular polarization, which can be shifted by varying the optical phases $\phi_L$ and $\phi_R$ (adapted after \cite{Robens:2016pol}). Two Raman laser beams perform transverse sideband cooling. A spiral phase plate (SPP) generates a hollow-tube laser beam (intensity profile in the inset), enhancing the transverse confinement of the atoms.
    Cooling along the longitudinal and transverse directions is evidenced by the suppressed blue detuned sidebands in the (b) microwave and Raman (c) sideband spectra, respectively.
    }
\end{figure}

The setup used for our experimental demonstration is schematically depicted in Fig.~\figref{fig4}{a}.
We start with a handful of Cs atoms, which are sparsely loaded in a one-dimensional polarization-synthesized optical lattice \cite{Robens:2016pol}.
Using the atom sorting technique presented in Ref.~\cite{Robens:2017}, a pair of atoms is then selected and repositioned to a relative distance of twenty lattice sites with a success rate of about $\SI{99}{\percent}$, mainly limited by an incorrect detection of the initial distance between the two atoms \cite{Alberti2015}.

To make the two atoms identical, we cool them to the ground state of the lattice site potential in which they are respectively trapped.
For this purpose, we use resolved sideband cooling, where
the sideband transitions are driven by microwave radiation \cite{Belmechri:2013} for the direction along the lattice axis and by two Raman lasers \cite{Kaufman:2012} for the transverse directions.
The optical lattice provides a tight confinement ($\nu_\parallel\approx \SI{120}{\kilo\hertz}$ sideband) along its longitudinal direction, while a hollow-tube potential collinear with the lattice axis also provides a tight confinement ($\nu_\perp \approx \SI{20}{\kilo\hertz}$ sideband) in the transverse directions.
These trap frequencies are much larger than the recoil frequency ($\approx \SI{2}{\kilo\hertz}$), thus ensuring that the Lamb-Dicke condition  necessary for ground state cooling is fulfilled.
We alternate between microwave and Raman sideband cooling 3 times in order to cool the atoms in both the longitudinal and transverse directions.
By allowing a slight ellipticity of the transverse potential, we lift the degeneracy of the transverse motional states, allowing Raman sideband cooling to be effective along both transverse directions.
At the end of the cooling process, the atoms are polarized in state $\ket{+}=\ket{F=4,m_F=4}$ with a probability of $\SI{99}{\percent}$.
Note that in this section $\ket{+}$ refers to a different Zeeman state than the one considered in the rest of this work (cf.\ Sec.~\ref{sec:wiring_qnt_circuits}).
With the $m_F$ state chosen for the experimental demonstration, it must be taken into account that the state-dependent potentials depend on both left and right circular polarization components of the trapping light field \cite{Lam:2021}.

Figures~\figref{fig4}{b} and \figref{fig4}{c} report a typical microwave and Raman sideband spectrum recorded after the cooling procedure, demonstrating a pronounced suppression of the cooling (blue detuned) sideband with respect to the heating (red detuned) sideband.
From the ratio of the sideband amplitudes \cite{Diedrich:1989}, we derive a ground state probability of $\approx \SI{95}{\percent}$ for the longitudinal direction and of $\approx \SI{84}{\percent}$ for each transverse direction.
Thus, the overall probability of occupying the motional ground state can be estimated as $\mathcal{P}_\text{3D} \approx \SI{95}{\percent} \times \SI{84}{\percent} \times \SI{84}{\percent} \approx \SI{67\pm 9}{\percent}$.

After sideband cooling, a magnetic field gradient along the lattice direction (\SI{11.6}{\gauss\per\centi\meter}) is ramped up in \SI{10}{\milli\second} and maintained until detecting the atoms by fluorescence imaging.
The magnetic field gradient induces a position-dependent Zeeman shift ($\approx \SI{1.2}{\kilo\hertz}$ per lattice site), which is used to selectively transfer one of the two atoms to the $\ket{-}$ state.
We perform such a selective spin flip by addressing the target atom with a microwave narrow-bandwidth $\pi$ pulse (Gaussian shape, \SI{7}{\kilo\hertz} rms width).
Spin-flip errors are clearly visible in the final fluorescence image, allowing them to be removed by post-selection.
The atom thus selected is then adiabatically transported in \SI{1}{\milli\second} to the site of the second atom by shifting the $V_-(x)$ lattice potential.
With the two atoms occupying the same site, we apply a fast microwave $\pi/2$ pulse (square shape, \SI{4.8}{\micro\second} duration).
This pulse acts much like the beam splitter of a Hong-Ou-Mandel optical interferometer, erasing the which-way information of the two impinging particles.
Last, we shift the $V_+(x)$ lattice potential to map the internal states, $\ket{+}$ and $\ket{-}$, to two different locations 10 lattice sites apart, where the atoms are detected by position-resolved fluorescence imaging.

Two identical atoms are expected to bunch together in the same lattice site with unit probability because of the Hong-Ou-Mandel interference (bosonic quantum statistics).
In practice, however, the two atoms can differ from each other because of their motional states.
For two atoms in orthogonal states (i.e., fully distinguishable particles), the outcomes resemble those obtained from the toss of two independent coins, yielding a bunching probability of $1/2$ (classical statistics).
For partially distinguishable atoms like ours, the probability to bunch is determined by the so-called quantum purity of the state, $\gamma \approx \mathcal{P}_\text{3D}^2$, which represents the probability of the two atoms to be indistinguishable: $\mathcal{P}_\text{bunch} = \gamma + (1-\gamma)/2$.
Thus, the Hong-Ou-Mandel interference is established if we can show that the bunching probability of the two atoms fulfills $\mathcal{P}_\text{bunch}>1/2$.

\begin{figure}[t]
    \includegraphics[width=\columnwidth]{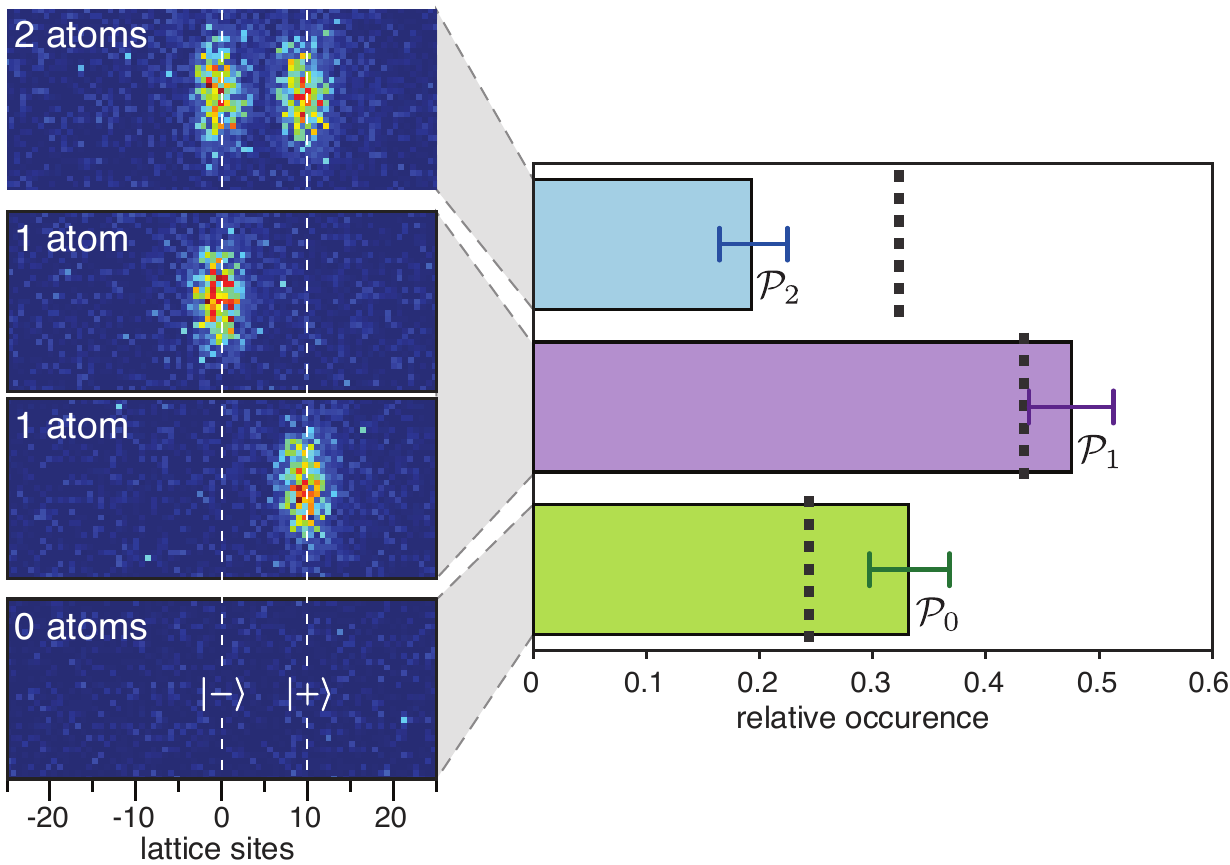}
    \caption{\label{fig3}
    \textbf{Detection of the Hong-Ou-Mandel interference.}
        Right: measured probability for the detection of zero ($\mathcal{P}_0$), one ($\mathcal{P}_1$), or two atoms ($\mathcal{P}_2$).
    The dashed lines show the case of distinguishable particles as a reference (see Appendix \ref{sec:monte_carlo_HOM}).
            The error bars represent the \SI{68}{\percent} Clopper-Pearson confidence intervals.
    Left: representative fluorescence images of the two atoms after the Hong-Ou-Mandel sequence, with the internal states $\ket{-}$ and $\ket{+}$ mapped to the sites $0$ and $10$.
    When the atoms are bunched in the same site, inelastic light-induced collisions result in either one or zero atoms being detected.
        Super-resolution microscopy allows resolving the lattice sites despite the diffraction-limited optical resolution of $\approx 4$ sites \cite{Alberti2015}. 
    }
\end{figure}

In the experiment, we distinguish three outcomes corresponding to the detection of zero, one, and two atoms in the final fluorescence image.
Figure~\ref{fig3} shows the experimentally recorded probability for each of them.
When both atoms are positively detected, the atoms are found at different locations, 10 sites apart (see fluorescence image in Fig.~\ref{fig3}).
Importantly, such an outcome can only occur when the two atoms have not bunched together.
Its probability is thus directly related to $\mathcal{P}_\text{bunch}$ through the expression: $\mathcal{P}_2 = S^2(1-\mathcal{P}_\text{bunch})$, where $S$ is the single-atom survival probability.
From the measurement of $\mathcal{P}_2 = \SI{19 \pm 3}{\percent}$ and the independent characterization of the single-atom survival probability $S=\SI{84\pm 1}{\percent}$ (see Appendix~\ref{sec:monte_carlo_HOM}), we therefore obtain $\mathcal{P}_\text{bunch} = \SI{73 \pm 4}{\percent}$, which exhibits a 5-$\sigma$ deviation from the reference value $1/2$.

Such a value of $\mathcal{P}_\text{bunch}$ establishes that the Hong-Ou-Mandel interference of the two atoms occurs with a probability $\gamma = 2\mathcal{P}_\text{bunch}-1 = \SI{45\pm8}{\percent}$, in good agreement with the value expected from the independent measurement of the ground state fraction, $\gamma \approx P_{3D}^2 \approx \SI{45\pm 12}{\percent}$.
We expect that $\gamma$ can be significantly improved in the future with more efficient ground-state cooling of the atoms in a three-dimensional optical lattice.
An analysis of all experimental outcomes, including those with zero and one atom detected, yields a value of $\mathcal{P}_\text{bunch} = \SI{73 \pm 6}{\percent}$ that is statistically consistent with the value we have derived from $\mathcal{P}_2$ only (see Appendix~\ref{sec:monte_carlo_HOM}).

In the future, it will be interesting to test not only the degree of indistinguishability between atoms, but also their degree of entanglement.
For that, one could study cross-correlations with a two-atom interferometer as presented in Refs.~\cite{Kofler:2012,LewisSwan:2015,Dussarrat:2017,Roos:2017}.

\section{Conclusions}

In this work, we have presented a scheme for the realization of programmable NISQ circuits with neutral atoms in state-dependent optical lattices.
Quantum circuits based on the proposed scheme can be easily reprogrammed and scaled up to hundreds of modes.
Both repogrammability and scalability are key to realize large random unitaries \textemdash a prerequisite for any large-scale boson sampling machine.
Furthermore, we have experimentally demonstrated the basic building block of an atom boson sampling machine by executing a quantum circuit with four modes and two indistinguishable atoms.
We observed the atoms bunching in pairs, thus revealing their bosonic nature.
The Hong-Ou-Mandel interference signal of the atom bunching is found to 
deviate from the outcome predicted for distinguishable (i.e., classical) particles by 5\,$\sigma$.
The degree of indistinguishability of the atoms is determined by the probability of occupying the motional ground state in the potential well of an optical lattice site.
We independently measured the motional ground state occupancy of the atoms and showed that it is in good agreement with the ground state occupancy inferred from the observed bunching probability.

We have discussed in detail how to wire quantum circuits using a one-dimensional state-dependent optical lattice.
Our analysis of NISQ devices has shown that controlling more than $M=500$ lattice sites will be required to reach a quantum advantage over best supercomputers.
Controlling such a large number of lattice sites may become difficult to realize in a one-dimensional geometry.
However, our scheme can be readily extended to two-dimensional state-dependent optical lattices~\cite{Groh:2016}, leading to a more compact and less resource-intensive platform. 

For future studies, it will be interesting to investigate the role of controlled coherent collisions among atoms \cite{Chin:2010}, which can be exploited to imprint collisional phases onto the quantum state when two or more particles meet at the same site \cite{Mandel04}.
The inclusion of such nonlinearities is shown to augment the amount of correlations in the output distribution \cite{Brunner:2018}.
For this reason, there is a potential that simulating a nonlinear boson sampler with classical computers will be an even harder task than the original linear problem.

Finally, we emphasize that beyond the boson sampling application, the proposed scheme can be used to implement reprogrammable parametrized quantum circuits \textemdash a key component for quantum machine learning \cite{Biamonte:2017,Benedetti:2019}.

After the completion of our manuscript, the first demonstration of an atomic boson sampler has been put forward~\cite{Young:2023}, involving 180 atoms spread across 1000 optical tweezer sites.

\appendix

\makeatletter
\renewcommand{\@hangfrom@section}[3]{#1\@if@empty {#2}{#3}{\MakeTextUppercase{#2}\@if@empty {#3}{}{:\ \MakeTextUppercase{#3}}}}
\makeatother

\section{Quantization axis}\label{app:q_axis_tilt}

The propagation direction of the addressing beam must be perpendicular to the optical lattice itself in order to allow it to be tightly focused on the individual sites of the lattice itself.
The only way to satisfy this geometric condition and the condition stated in Sec.~\ref{sec:wiring_qnt_circuits} that the quantization axis must have a nonzero component along the addressing beam is to have the quantization axis form an angle $\alpha\neq 0$ with the direction of the optical lattice.

However, tilting the quantization axis by $\alpha$ with respect to the optical lattice laser beams causes a wobbling of the optical lattice potential depth, $V^0_\pm$, during the state-dependent shift operations.
Because of this wobbling, the lattice depth is reduced by a factor $|\cos(\alpha)|$ during transport.
The extent of this effect is very small for small angles.
For example, for $\alpha=\SI{15}{\degree}$ the lattice depth is reduced by only $\approx \SI{3}{\percent}$.
This wobbling can be easily taken into account and compensated for in the design of transport operations by optimal control \cite{Lam:2021}.

\section{Pair distribution and extended model}\label{app:Pair}
In the limit of large $N$ and after averaging over random quantum circuits, the state of the $N$ bosons is described by a uniform statistical mixture $\rho_u$, where all possible configurations of the $N$ bosons over the $M$ modes are equally weighted \cite{Arkhipov:2012}.
Using this result, we derive a general expression $P(k_2,k_3)$ to account for the probability of finding exactly $k_2$ sites occupied by a pair and $k_3$ sites occupied by a triplet, whereas the other sites are singly occupied.

To calculate $P(k_2,k_3)$, we first determine the number of configurations containing $k_3$ sites occupied by a triplet of atoms.
For that, we consider that there are $\binom{M/2}{k_3}$ possible combinations to arrange $k_3$ triplets in $M/2$ sites, where we assume for simplicity that the overall number of modes, $M$, is even.
Given the two spin states, for each site occupied by a triplet, we have $\big(\hspace{-2.5pt}\binom{2}{3} \hspace{-2.5pt}\big)=4$ possible spin configurations, where the double brackets denote the multiset coefficient.
Thus, the previous number of combinations must be multiplied by $4^{k_3}$ to obtain the total number of configurations for the $k_3$ triplets.
Next, we consider that there are $\binom{M/2-k_3}{k_2}$ combinations to arrange $k_2$ pairs in the remaining $M/2-k3$ sites.
This number must be multiplied by $3^{k_2}$ to account for the $\big(\hspace{-2.5pt}\binom{2}{2} \hspace{-2.5pt}\big)=3$ different spin configurations for each pair.
Now, there are $\binom{M/2-k_2-k_3}{N-2k_2-3k_3}$ different combinations to place the remaining $N-2k_2-3k_3$ particles in the rest of the sites.
For each singly occupied site, there are only two possible spin configurations.
Finally, to obtain the probability $P(k_2,k_3)$, all the configurations should be divided by the overall number of possible bosonic configurations, which is given by the multiset coefficient $\big(\hspace{-2.5pt}\binom{M}{N} \hspace{-2.5pt}\big)$.
Hence, the probability of having exactly $k_2$ sites occupied by pairs and $k_3$ sites occupied by triplets is given by 
\begin{eqnarray}\label{NPairProb}
P(k_2,k_3)&=&4^{k_3}\binom{M/2}{k_3}3^{k_2}\binom{M/2-k_3}{k_2}2^{N-2k_2-3k_3}\nonumber\\ &\times&\binom{M/2-k_2-k_3}{N-2k_2-3k_3}\Big/\bigg(\hspace{-4.5pt}\binom{M}{N} \hspace{-4.5pt}\bigg).
\end{eqnarray}

\begin{figure}[t]
    \includegraphics[width=0.9\columnwidth]{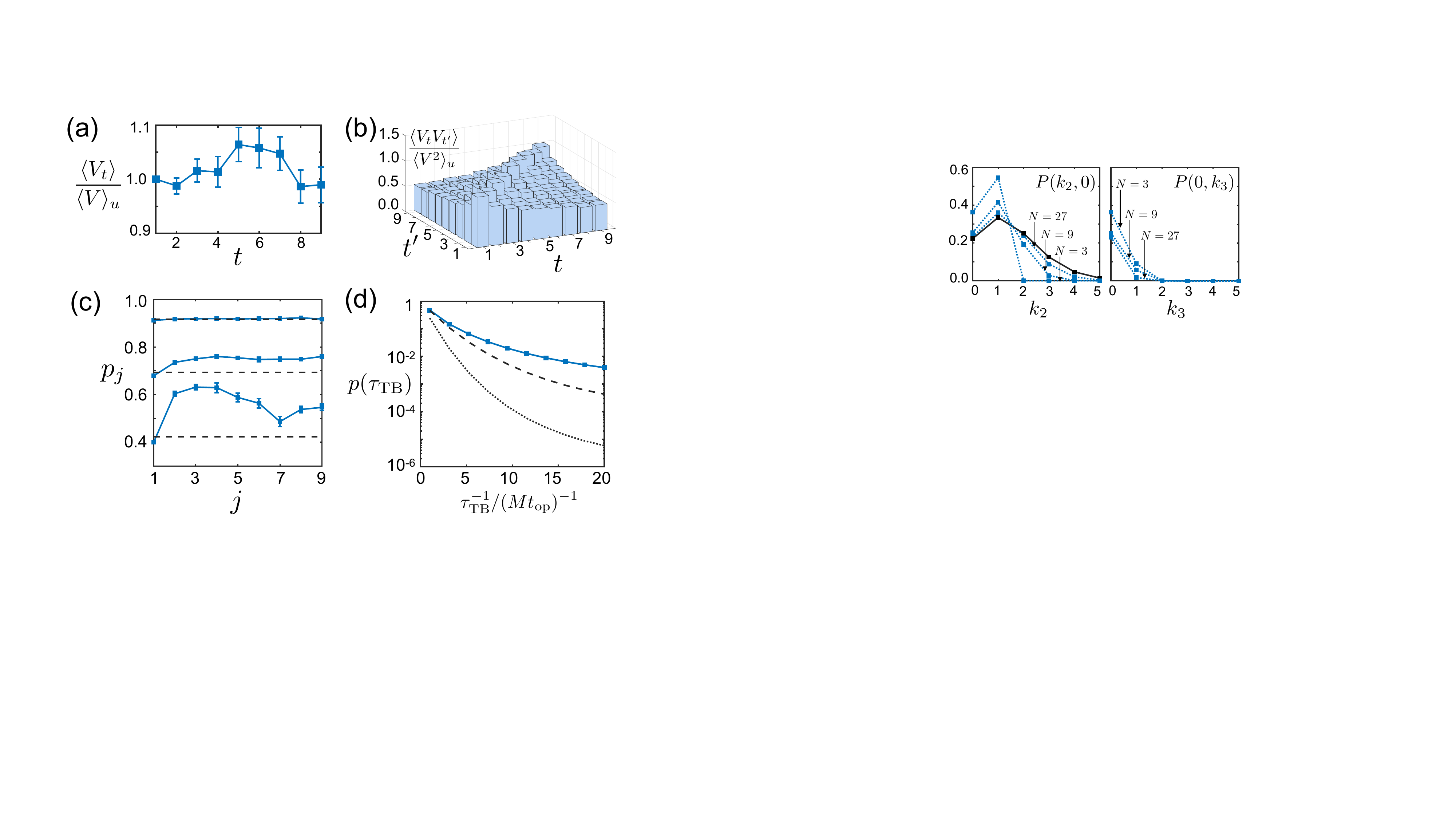}
    \caption{\label{fig_pairprob}
    \textbf{Extended probability distribution of particle pairs and triplets.}
    Left: $P(k_2,0)$ versus $k_2$ is plotted (dotted lines) for $N=3, 9$ and $27$ and $c=1$. For increasing $N$, $P(k_2,0)$ tends to Eq.~(\ref{eq:SitePoissonian}) (solid line).
    Right: $P(0,k_3)$ versus $k_3$ is plotted (dotted lines) for $N=3, 9$ and $27$ and $c=1$. For increasing $N$, $P(0,k_3>0)$ tends to zero.}
\end{figure}

In the limit of large $N$, the expression in Eq.~(\ref{NPairProb}) yields a vanishing probability to have $k_3>0$ triplets, $\lim_{N\to\infty} P(k_2,k_3>0)=0$.
Instead, when $k_3=0$, Eq.~(\ref{NPairProb}) converges to a Poisson distribution for the occurrence of $k_2$ pairs,
\begin{equation}\label{eq:SitePoissonian}
\lim_{N\to\infty} P(k_2,0)=\frac{\lambda^{k_2}}{e^{\lambda}k_2!},
\end{equation}
with average value $\lambda=3/(2c)$. Here, the constant factor $c$ denotes the ratio $c=M/N^2$, which in the main text has simply been assumed equal to 1.
Figure~\ref{fig_pairprob} shows the distributions $P(k_2,0)$ (left panel) and $P(0,k_3)$ (right panel) for an increasing number of particles $N=3, 9$ and 27, assuming $c=1$.
The graphs show that the probability associated with the collision-free subspace, $P(0,0)$, tends to the finite value $\exp[-3/(2c)]$.
We also notice that for a relatively small number of particles $N\lesssim 10$, the probability of finding a triplet, $P(0,1)$, has a nonnegligible value between $0.05$ and $0.1$.

It is straightforward to generalize Eq.~(\ref{NPairProb}) to higher occupations (quartets, quintets, etc).
However, the probability of having more than 3 particles in a site is negligible even for small number of particles and can thus be safely neglected. 

The extended distribution $P(k_2,k_3)$ can be used to estimate the effect of two-body losses on the uniform state $\rho_u$.
Two-body losses are assumed to reduce the survival probability of a state containing $k_2$ pairs by a factor $e^{-k_2 t/\tau_{\rm TB}}$.
A particle triplet decays three times faster than a pair because there are 3 possible combinations for the colliding pairs.
This can also be understood through the action of the particle decay operator $V(t)$ (defined in Appendix~\ref{App:ExactSim}) on a state with $k_3$ particle triplets, where $V(t)|k_3\rangle\langle k_3|V^\dagger(t)=\exp{(-3k_3t/\tau_{\rm TB})} |k_3\rangle\langle k_3|$.
The total survival probability is thus obtained evaluating the sum 
\begin{equation}\label{SurvProbExt}
P_{\rm step}^\text{TB}=\sum_{k_3=0}^{N/3} \sum_{k_2=0}^{N/2-k3}P(k_2,k_3) e^{-\frac{(k_2+3k_3)t_\text{step}}{\tau_{\rm TB}}}.
\end{equation}
The sum in Eq.~(\ref{SurvProbExt}) can be simplified in the limit of large $N$ by ignoring particle triplets. Using the approximate expression in Eq.~(\ref{eq:SitePoissonian}) for $P(k_2,0)$, we obtain
\begin{equation}
\label{SurvProb1_app}
P_{\rm step}^\text{TB} \approx \hspace{-1pt} \sum_{k_2=0}^{N/2} P(k_2,0) e^{-\frac{k_2 t_\text{step}}{\tau_{\rm TB}}} \approx
 \exp{\left[\frac{3}{2c}\hspace{-1pt}\left(e^{-\frac{t_{\rm step}}{\tau_{\rm TB}}}-1\right)\right]},
\end{equation}
which corresponds to Eq.~(\ref{SurvProb1}) in the main text when $c=1$.
The expression on the right-hand side of Eq.~(\ref{SurvProb1_app}) is obtained by computing the limit as $N$ tends to infinity.

Note that in Fig.~\ref{fig2} and in Appendix~\ref{App:ExactSim}, the extended model of Eq.~(\ref{SurvProbExt}) is used instead of Eq.~(\ref{SurvProb1}) to describe the two-body losses, because it also applies in the limit of a small number of particles.

\section{Scaling law for photonic and classical boson samplers}\label{app:OtherPlat}

Along the lines of Ref.~\cite{Wang-Hui:2017}, the sampling rate of a photonic boson sampler is given by 
\begin{equation}\label{photonscaling}
R_{\rm ph}=\frac{1}{e}\frac{R_0}{N}\eta^{N}, 
\end{equation}
where $R_0/N$ is the rate in which $N$ indistinguishable photons are created and $\eta=\eta_{f}\eta_{c}^d$ is the success probability of a single photon going through state preparation, circuit transmission, and successful detection. 
The success probability $\eta$ is a product of a fixed probability $\eta_{f}$, which does not scale with the number of modes in the circuit and depends only on the state preparation and detection, and the circuit transmission probability $\eta_{c}^M$, which accounts for the chance that a photon is absorbed at each beam splitter.
For a square circuit \cite{Clements:2016} with $M=N^2$, the sampling rate $R_{\rm ph}$ is mainly determined by the term $\eta_{c}^{N\hspace{0.1pt}M}$ for sufficiently large $N$.

Our conservative values (lower edge of the purple area in Fig.~\ref{fig2}) are based on Ref.~\cite{Wang:2019e}. 
The authors use a single-photon source working at $R_0=76$ MHz and report a boson sampling rate of $R_{\rm ph}=295$\,Hz for 5-photons in a $60{\times}60$ optical circuit. 
The transmission probability through the entire circuit is reported as $98.7\%$ and correspondingly $\eta_{c}=0.987^{1/60}$, which in turn allows us to extract the fixed preparation and detection probability of $\eta_{f}=0.14$.
For the state-of-the-art estimate (upper edge of the purple area in Fig.~\ref{fig2}), we assume an overall increased fixed preparation and detection probability of $\eta_{f}=0.65$~\cite{Wang:2018}.

For a classical boson sampler, the time required by a recently developed algorithm~\cite{Neville:2017} based on Metropolised independent sampling to generate a valid sample scales as 
\begin{equation}\label{computerrate}
R_{\rm cl}=2^{-N}/(100\,\tilde{a}\,N^2)
\end{equation}
where $\tilde{a}$ relates to the speed of the classical computer. This value has been reported to be $\tilde{a}=\SI{3e-15}{\second}$ in the case of the Tianhe-2 supercomputer~\cite{Wu:2018}. For the case of an ordinary computer, we choose this value to be $\tilde{a}=\SI{3e-9}{\second}$.

\section{Benchmark with exact simulations}\label{App:ExactSim}

We derive a master equation model in order to benchmark the sampling rate formula in Eq.~(\ref{AtomicScaling}).
In our model, an $N$-particle initial state is given by $|\psi_0\rangle$, where $\langle\psi_0|\psi_0\rangle=1$ and $\sum_{m=1}^{M}{n}_m|\psi_0\rangle=N|\psi_0\rangle$. Here, ${n}_m$ is the number operator acting on mode $m$. The evolution of such a quantum state is given by
\begin{equation}\label{state}
|\psi\rangle= U_{M}V( t_{\rm step})U_{M-1}\dots U_{1}V(t_{\rm step})|\psi_0\rangle,
\end{equation}
where $U_j$ are the coherent operations at step $j$, and $V(t)=\exp{(-Ht)}$ represents particle decay operator acting for a time $t$.
The Hermitian operator $H$ is
\begin{equation}\label{DecayHamil}
H=\frac{1}{2\tau_{\rm BG}}\sum_{s=1}^{M/2}{n}_s + \frac{1}{4\tau_{\rm TB}}\sum_{s=1}^{M/2}{n}_s({n}_s-1),
\end{equation}
where ${n}_s$ is the number operator counting the number of atoms in the lattice site $s$.
In Eq.~(\ref{DecayHamil}), the first and second terms account for single-body and two-body decay processes, respectively.
The effect of $V(t_{\rm step})$ is to leave the state in the $N$-particle subspace (i.e., the collision-free subspace in which we are interested), but with a reduced norm.
When the operator is applied to a state with $k$ pairs, it reduces the norm by a factor $\exp{(-N t_{\rm step}/2\tau_{\rm BG})}\exp{(-k\hspace{0.3pt} t_{\rm step}/2\tau_{\rm TB})}$.
The survival probability of the boson sampler is thus characterized by the norm squared, $\langle\psi |\psi\rangle$.

\begin{figure}[t]
    \includegraphics[width=\columnwidth]{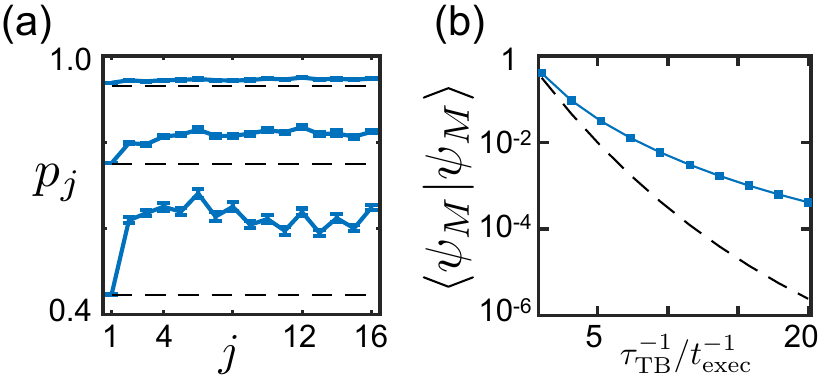}
    \caption{\label{fig_exactsim}
	\textbf{Numerical simulation benchmark of the survival probability.} Blue lines represent the exact simulation for $N=4$ and $M=16$, where each point is the average of 30 realizations with different random unitaries.
	(a) Success probability $p_j$ from time step $j-1$ to $j$ for different decay rates $\tau_{\rm TB}=t_{\rm exec}$, $\tau_{\rm TB}=t_{\rm exec}/5$ and $\tau_{\rm TB}=t_{\rm exec}/20$.
	The black dashed lines are obtained using the simple model in Eq.~(\ref{SurvProbExt}).
	(b) Total survival probability $\langle \psi_M |\psi_M\rangle$ versus $\tau_{\rm TB}^{-1}$.
	The black dashed line represents the simple model, $(P_{\rm step}^{\rm TB})^M$.}
\end{figure}

We use the master equation model given in Eq.~(\ref{state}) to carry out numerical simulations for up to $N=4$ bosons, $M = N^2$ modes, and $M$ steps.
The exact numerical simulations allow us to validate the survival probability model in Eq.~(\ref{SurvProbExt}), which was used to estimate the sampling rate of an atom boson sampler in Sec.~\ref{ScalingAndErrors}.
We consider in particular the survival probability from time step $j{-}1$ to $j$, which is given by $p_j=\langle\psi_j|\psi_j\rangle/\langle\psi_{j-1}|\psi_{j-1}\rangle$, where $|\psi_j\rangle=\prod_{i=1}^{j}U_iV(t_{\rm step})|\psi_0\rangle$.
In Fig.~\figref{fig_exactsim}{a}, we plot $p_j$ versus $j$ and for different values of $\tau_{\rm TB}$.
For clarity, we omit the effect of single-particle losses as their effect is trivial, and for simplicity we choose as initial state a pure state given by a uniform superposition of all bosonic states of $N$ particles.
For $\tau_{\rm TB}=t_{\rm exec}$, we find an almost exact correspondence between $P_{\rm step}$ (black dashed line) and $p_j$ (blue solid curve) for all steps $j$, suggesting that, for weak losses, Eq.~(\ref{SurvProbExt}) provides an accurate description of the overall survival probability.
For lower values of $\tau_{\rm TB}$, such as $\tau_{\rm TB}=t_{\rm exec}/5$ or $\tau_{\rm TB}=t_{\rm exec}/20$, this correspondence only holds for the initial state, while for the rest of the steps, $P_{\rm step}^{\rm TB}<p_j$, showing that Eq.~(\ref{SurvProbExt}) represents a lower bound on the survival probability for each step.
In Fig.~\figref{fig_exactsim}{b}, we evaluate the total survival probability  $\langle \psi_M |\psi_M\rangle$ as a function of $\tau_{\rm TB}$ (blue solid curve), and compare it with the result from the simple model, $(P_{\rm step}^{\rm TB})^M$.
We find that the simple model gives a correct description of two-body losses when $\tau_{\rm TB}\gtrsim t_{\rm exec}$ and a lower bound estimate when $\tau_{\rm TB} \lesssim t_{\rm exec}$.
Although numerical benchmarks have been performed for up to $N=4$, to our knowledge, it is reasonable to assume that the observed results are generalizable to a much larger number of particles.

\section{Monte Carlo analysis of the Hong-Ou-Mandel experiment}
\label{sec:monte_carlo_HOM}

In order to extract $\mathcal{P}_\text{bunch}$ from the experimental results, it is important to consider that atoms bunched in the same site cannot be directly detected by fluorescence imaging without losing them.
In fact, independent measurements performed in our apparatus on bunched atoms show that with a probability $P_\text{LIC,0} = \SI{71 \pm 5}{\percent}$ no atom is left, while with a probability $\mathcal{P}_\text{LIC,1}=1-\mathcal{P}_\text{LIC,0}$ just a single atom is detected in the final fluorescence image as a consequence of light-induced collisions \cite{Schlosser:2001}.
Hence, identical atoms lead to the detection of either zero or one atom, while distinguishable atoms are both detected if after the  Hong-Ou-Mandel $\pi/2$ pulse they occupy different internal states.
Examples of the three possible outcomes with $0$, $1$ and $2$ atoms detected are shown in Fig.~\figref{fig3}{a}.

To measure the single-atom survival probability $S$, we performed independent measurements similar to the Hong-Ou-Mandel interference experiment outlined in the main text, omitting the microwave $\pi/2$ pulse.
Without the $\pi/2$ pulse, $\mathcal{P}_\text{bunch}$=0, independent of the quantum purity of the state, $\gamma$.
From this calibration experiment, we can directly determine $S=\SI{84\pm 1}{\percent}$.
The value of $S$ is limited by losses at the beginning of the transverse cooling process, which could be avoided in the future with an improved experimental procedure.

We employ a Monte Carlo analysis to more rigorously analyze the statistical outcomes recorded for the Hong-Ou-Mandel experimental sequence that is described in Sec.~\ref{HOMExpSect}.
The Monte Carlo simulation uses the predetermined light-induced collision probability $\mathcal{P}_\text{LIC,0}$ and the single-atom survival probability $S$. 
Further input parameters are the bunching probability $\mathcal{P}_\text{bunch}$, 
the addressing probability of the narrow bandwidth MW pulse, and the probability to successfully reconstruct the position of the atoms.
Using a nonlinear least squares regression, we fit the generated Monte Carlo events to the measured data to extract the underlying experimental parameters, yielding $\mathcal{P}_\text{bunch} = \SI{73 \pm 6}{\percent}$.

\vspace{3mm}
\begin{acknowledgments}
We acknowledge financial support from the NRW-Nachwuchsforschergruppe ``Quantenkontrolle auf der Nanoskala'', the Deutsche Forschungsgemeinschaft SFB project OSCAR, the Basque Government with PhD grant PRE-2015-1-0394, the Junta de Andaluc\'ia (grants P20-00617 and US-1380840), the Spanish Ministry of Science, Innovation, and Universities (grants PID2019-104002GB-C21 and PID2019-104002GB-C22), the National Natural Science Foundation of China (grant 12075145), and Science and Technology Commission of Shanghai Municipality (grant 2019SHZDZX01-ZX04).
A.\,A acknowledges support from the Alexander von Humboldt Foundation, I.\,A.\ from the European Union’s Research and Innovation Programm (grant SuperQuLAN 899354), C.\,R.\ from the Studienstiftung des deutschen Volkes (grant 421987027).

C.\,R.\ and I.\,A.\ contributed equally to this work.
\end{acknowledgments}

\bibliographystyle{apsrev4-2}

\end{document}